\documentclass[12pt,noshowpacs,nofootinbib,notitlepage,amsmath]{revtex4-1}
\allowdisplaybreaks
\usepackage[colorlinks=true,citecolor=blue,linkcolor=blue,urlcolor=blue]
{hyperref}
\usepackage{graphicx,color}
\usepackage{multirow}
\usepackage{enumerate}

\DeclareMathOperator{\Lapl}{\mathcal{L}}
\newcommand{\Ltran}{\widehat{\psi}}
\newcommand{\Rop}{\mathcal{R}}
\newcommand{\TI}{\mathcal{I}}
\newcommand{\TS}{\mathcal{S}}

\begin{document}  
\title{\Large Gravitational Wave Perturbations on a Kerr Background and Applications for Echoes}

\author{Randy S. Conklin}
\email{rconklin@physics.utoronto.ca}
\affiliation{Department of Physics, University of Toronto, Toronto, Ontario, Canada,  M5S 1A7}
\begin{abstract}
Gravitational wave echoes may imply physics beyond general relativity; however, several approaches to searching for echoes require fitting the data to toy templates which are not physically motivated. Here, I create a procedure which outputs a more realistic form for echoes, starting from initial conditions, by calculating the modulated transfer function.\end{abstract}
\maketitle

\section{Introduction}\label{Introduction}
The theory supporting echoes relies on the possibility that the traditional idea of black holes may be false. Normally, black holes are considered to be ultimately absorbing entities, but if black holes contain any type of inner reflective surfaces, which we may call ``walls,'' or if they are in reality exotic compact objects which do not feature horizons, then it is possible that what would otherwise be terminally infalling objects could instead escape to infinity. Such walls can arise from various modifications of general relativity, some examples of which are theories where quantum effects or strong gravity take over within a Planck distance of where the horizon would otherwise be~\cite{Original}.\footnote{For the sake of brevity and simplicity, I will be liberal in my usage of the term ``black hole,'' ascribing it sometimes to other exotic objects, or to ``black hole'' counterparts in theories other than GR.}

To explain the origin of echoes, I will first describe the picture for a Schwarzschild background, save that it may feature an exotic wall at some location near the would-be horizon. In this case, one can perturb the metric and solve Einstein's equations to obtain a wave equation with an effective potential~\cite{BlackBranes}.\footnote{This works for scalar, vector, and gravitational perturbations.} The potential is an angular momentum barrier that peaks near the light ring radius of $r=3M$ and goes to zero for $r$ approaching the horizon or infinity. If a wall is present sufficiently far from $3M$, then a cavity exists between the wall and the barrier. If the wall is reflecting, then perturbations may bounce back and forth within the cavity. Since the barrier is finite, a fraction of the perturbation will leak out with each cycle, and pulses will be emitted at regular intervals. 

In the simplest case, a distant observer then detects ``echoes'' as periodic pulses separated by a time delay related to the distance between the wall and the barrier. The frequency content of these echoes is related to the partially trapped modes of the cavity, and features resonances, called quasi-normal modes, which can be classified by the locations in the complex plane of their corresponding poles, and by orbital numbers. These resonances specify the mass---and spin, if the background is rotating---of the geometry and the nature of the wall. Their spacing, which is nearly constant throughout the spectrum and proportional to the inverse of the time delay, determines the location of the wall. Thus, the frequency content and time delay of echoes encode the characteristics and geometry of the underlying perturbed object, and may distinguish between different theories. 

Searches for echoes have for the most part assumed a toy model template for echo signals and matched this onto the data~\cite{Abyss,Abyss2,LowSignificance,Rico}. Given its preliminary nature, this template is, naturally, somewhat physically unmotivated, and ignores certain traits of the potential which are now known to significantly alter the signals.\footnote{One example is that the potential acts as a high bandpass filter and leads to a sequentially lower frequency content for each subsequent pulse.} This presents a problem since most of these search methods can be extremely sensitive to details and require an accurate template, but the template has known difficulties. 

In~\cite{Windows}, a method was introduced which extracts information from the primary features of echoes while being less sensitive to auxiliary details. There, the focus was on the transfer function, and the search target was the spectrum of nearly evenly separated resonances characteristic of echoes. The key equation is the following, which expresses the spectrum of echoes observed by a distant detector:
\begin{align}
\psi_\omega(x) = e^{i\omega x}{\cal K(\omega)}\int_{-\infty}^\infty p(x^\prime,\omega)\psi_\textrm{left}(x^\prime,\omega)I(x^\prime,\omega) dx',
\label{eq:obspsi}
\end{align}

For the present purposes, the two most salient features of (\ref{eq:obspsi}) are 1) The transfer function ${\cal K}$ and 2) The integral part, which I will refer to as the source integral or the modulation integral. Other details are explained in Sec.\ref{subsec:Modulation}, with further information available in~\cite{Windows}. Currently, little work has been done to study $I$\footnote{Progress has been made in~\cite{AnalyticalEchoes}, where the approximately analytical regime of the perturbation equation is explored along with implications for data analysis.}. This is especially true for the case of a gravitationally perturbed rotating background, and yet this is precisely the astrophysically relevant case. Given the importance of ${\cal K}$ as a search target, it is significant to understand more completely how (\ref{eq:obspsi}) is influenced by $I$.

In this paper, I calculate $I$ from initial conditions, and thus determine the modulated form of the transfer function and so the observable spectrum for echoes. 

The structure of this paper is as follows. In Section II, I begin with a partial differential equation describing gravitational perturbations on a spinning background, proceed to derive the source term of the equation from initial conditions, and convert the result to a numerically tractable form. In Section III, I exhibit results relevant for data searches, including sample modulations of the transfer function and a series of corresponding echoes.

\section{Deriving the Source Term $I$}\label{SectionOne}

We are interested in studying gravitational waves, for which we will use the Teukolsky master equation~\cite{Teuk}. For spin $s = -2$, this determines the dynamics of gravitational perturbations on a Kerr background of mass $M$ and spin $a$. In vacuum, this reads
\begin{equation}\label{eq:Teuk}
\begin{split}
&\left[\frac{(r^2+a^2)^2}{\Delta}-a^2\sin^2\theta\right] \frac{\partial^2\psi}{\partial t^2} + \frac{4Mar}{\Delta}\frac{\partial^2\psi}{\partial t\partial\phi} + \left[\frac{a^2}{\Delta}-\frac{1}{\sin^2\theta}\right]\frac{\partial^2\psi}{\partial\phi^2} - \Delta^{-s}\frac{\partial}{\partial r} \left( \Delta^{s+1}\frac{\partial\psi}{\partial r}\right)\\
&-\frac{1}{\sin\theta}\frac{\partial}{\partial\theta}\left(\sin\theta\frac{\partial\psi}{\partial\theta}\right) - 2s\left[\frac{a(r-M)}{\Delta}+\frac{i\cos\theta}{\sin^2\theta}\right]\frac{\partial\psi}{\partial\phi} - 2s\left[\frac{M(r^2-a^2)}{\Delta} - r - ia\cos\theta\right]\frac{\partial\psi}{\partial t}\\
& + (s^2\cot^2\theta-s)\psi = 0 ,
\end{split}
\end{equation}
where $\Delta=(r^2-2Mr+a^2)$. In its current form, (\ref{eq:Teuk}) is not very useful to us. One issue is that it does not reduce to the Regge-Wheeler equation for the spinless case. To enforce this, we will change variables to obtain the Sasaki-Nakamura (SN) equation~\cite{SN}. Additionally, the SN equation is asymptotically simple with a flat potential near the horizon and a zero potential at infinity.

To obtain the SN equation with the correct source term, we must 1) Separate variables to obtain a purely radial equation including initial conditions, 2) Transform the resulting equation into SN form, and 3) Solve for the source term in SN.

\subsection{Separation of Variables and Initial Conditions}

\subsubsection{Separation of Azimuthal and Time Dependence, and the Initial Conditions}

We first separate out the azimuthal dependence with the definition
\begin{equation}
\psi = \sum_m \psi_m(t,r,\theta)e^{im\phi}.
\end{equation}

Then we trade the time dependence for an equation in frequency space with initial conditions. We do so by applying the Laplace transform for a perturbation which turns on at some time $t_0$:
\begin{equation}
\Lapl \psi_m(t,r,\theta) \equiv \Ltran_m(\omega,r,\theta) = \int_{t_0}^{\infty}\psi_m(t,r,\theta)e^{i\omega t}dt .
\end{equation}
When applied onto time derivatives, this transform gives
\begin{equation}
\Lapl\left[ \frac{\partial \psi_m}{\partial t} \right] = -i\omega \Lapl \psi_m - e^{i\omega t_0} \psi_m(t_0,r,\theta)
\end{equation}
Thus, with the primes denoting time derivatives,
\begin{equation}
\begin{split}
\Lapl\left[ \frac{\partial^2 \psi_m}{\partial t^2} \right] =  &-i\omega \Lapl \psi_m^\prime - e^{i\omega t_0}\psi_m^\prime(t_0,r,\theta) \\
= &-i\omega (-i\omega \Lapl \psi_m - e^{i\omega t_0} \psi_m(t_0,r,\theta)) - e^{i\omega t_0}\psi_m^\prime(t_0,r,\theta)
\end{split}
\end{equation}
Or
\begin{equation}
\Lapl\left[ \frac{\partial^2 \psi_m}{\partial t^2} \right] = -\omega^2 \Lapl\psi_m + i\omega e^{i\omega t_0}\psi_m(t_0,r,\theta) - e^{i\omega t_0}\psi_m^\prime(t_0,r,\theta)
\end{equation}

We must use the Laplace transform instead of the Fourier transform since Fourier transforms, for causal functions, are well-defined for~\cite{Andersson}
\begin{equation}
\int_{t_0}^\infty |\psi_m|^2 dt < \infty,
\end{equation}
but this is not satisfied for functions which on average continue to grow in time. For a Kerr black hole, it is possible to use a Fourier transform, but, in our case, where we truncate the background and impose a perfectly reflecting boundary condition, for a patch of frequencies slightly below the BH rotational frequency, modes are reflected off the potential with a reflection rate greater than one.\footnote{It should be noted that the open question of which boundary condition to use is important, and our choice here is somewhat arbitrary.}

Another way to see this is to recognize that echo modes are quasinormal modes associated with complex frequencies. The imaginary part of such frequencies leads to damping or growth in time. For a truncated Kerr black hole with our boundary condition, some of these modes have imaginary parts which cause growth in time, hence the instability.

The Laplace transform is well-defined even for a truncated Kerr black hole. It is equivalent to the Fourier transform, but with $\omega$ complex such that its imaginary part counteracts the growth in $\psi$, and the integrand of the transform is always finite and approaches zero for time approaching infinity. 

Later, when we generate echoes in the time domain, we will use the inverse of the Laplace transform
\begin{equation}\label{eq:InverseLaplace}
\psi_m = \frac{1}{2\pi}\int_{-\infty + ic}^{\infty + ic}  \Ltran_m e^{-i\omega t} d\omega,
\end{equation}
where the value of $ic$ is constrained such that the integration contour is above, with respect to the imaginary axis, all poles of $ \Ltran_m$. This causes the initial transformation to be sufficiently damped and thus well-defined.

Laplace transforming the Teukolsky equation with $\phi$ separated out, the terms with time derivatives become
\begin{equation}
\begin{split}
&\left[\frac{(r^2+a^2)^2}{\Delta}-a^2\sin^2\theta\right] (-\omega^2 \Lapl\psi_m + i\omega e^{i\omega t_0}\psi_m(t_0,r,\theta) - e^{i\omega t_0}\psi_m^\prime(t_0,r,\theta))\\
& + \frac{4iMamr}{\Delta}(-i\omega \Lapl \psi_m - e^{i\omega t_0} \psi_m(t_0,r,\theta))\\
& - 2s\left[\frac{M(r^2-a^2)}{\Delta} - r - ia\cos\theta\right](-i\omega \Lapl \psi_m - e^{i\omega t_0} \psi_m(t_0,r,\theta)).
\end{split}
\end{equation}

The terms involving $t_0$ explicitly are initial conditions, and we bring these to the right hand side of the equation and refer to them as source terms. Suppressing subscripts for simplicity, the operator side of the Teukolsky master equation becomes
\begin{equation}\label{eq:SinSquared}
\begin{split}
&-\left[\frac{(r^2+a^2)^2}{\Delta}-a^2\sin^2\theta\right]\omega^2 \Ltran - \frac{4iMamr}{\Delta}i\omega \Ltran - m^2\left[\frac{a^2}{\Delta}-\frac{1}{\sin^2\theta}\right]\Ltran - \Delta^{-s}\frac{\partial}{\partial r} \left( \Delta^{s+1}\frac{\partial\Ltran}{\partial r}\right)\\
& - \frac{1}{\sin\theta}\frac{\partial}{\partial\theta}\left(\sin\theta\frac{\partial\Ltran}{\partial\theta}\right) - 2ims\left[\frac{a(r-M)}{\Delta}+\frac{i\cos\theta}{\sin^2\theta}\right]\Ltran + 2s\left[\frac{M(r^2-a^2)}{\Delta} - r - ia\cos\theta\right]i\omega \Ltran\\
& + (s^2\cot^2\theta-s)\Ltran .
\end{split}
\end{equation}
We can rearrange this to take the following simple form:
\begin{equation}
\Rop \Ltran + \Theta \Ltran + A \Ltran = \TI,
\end{equation}
where $\Rop$ is a purely radial operator, $\Theta$ is purely angular, and $A$ is an eigenvalue of $\Theta$. The source term with initial conditions, $\TI$, depends on $r$ and $\theta$.

Next we choose to exchange the $\sin^2\theta$ term in (\ref{eq:SinSquared}) for $1 - \cos^2\theta$ and to bring the constant term into the radial equation. We get
\begin{equation}
\begin{split}
\Rop = &\Delta^{-s}\frac{\partial}{\partial r} \left( \Delta^{s+1}\frac{\partial}{\partial r}\right) + \omega^2 \frac{(r^2+a^2)^2}{\Delta} - \frac{4Mamr}{\Delta}\omega + m^2\frac{a^2}{\Delta} + 2ims\frac{a(r-M)}{\Delta}\\
& - 2i\omega s\left(\frac{M(r^2-a^2)}{\Delta}-r\right) - a^2\omega^2 - A\\
\Theta = & a^2\omega^2\cos^2\theta - \frac{m^2}{\sin^2\theta} + \frac{1}{\sin\theta}\frac{\partial}{\partial\theta}\left(\sin\theta\frac{\partial}{\partial\theta}\right) - 2ms\frac{\cos\theta}{\sin^2\theta} - 2sa\omega\cos\theta - s^2\cot^2\theta + s\\
\TI = & \left[\frac{(r^2+a^2)^2}{\Delta}-a^2\sin^2\theta\right](i\omega e^{i\omega t_0}\psi_m(t_0,r,\theta) - e^{i\omega t_0}\psi_m^\prime(t_0,r,\theta))\\
& - \frac{4iMmar}{\Delta}e^{i\omega t_0} \psi_m(t_0,r,\theta) + 2s\left[\frac{M(r^2-a^2)}{\Delta} - r - ia\cos\theta\right]e^{i\omega t_0} \psi_m(t_0,r,\theta)
\end{split}
\end{equation}
Rearranging $\TI$, we can write
\begin{equation}\label{eq:InitialConditions}
\begin{split}
\TI = & - e^{i\omega t_0}\left(-i\omega\left[\frac{(r^2+a^2)^2}{\Delta}-a^2\sin^2\theta\right] + \frac{4iMmar}{\Delta} - 2s\left[\frac{M(r^2-a^2)}{\Delta} - r - ia\cos\theta\right]\right)u\\
& - e^{i\omega t_0}\left(\left[\frac{(r^2+a^2)^2}{\Delta}-a^2\sin^2\theta\right]\right)v ,
\end{split}
\end{equation}
where $u = \psi_m(t_0,r,\theta)$ and $v = \psi_m^\prime(t_0,r,\theta)$.

When deriving their equation, Sasaki and Nakamura adopt a slightly different notation for the radial operator. To obtain their form for the equation, we define a new eigenvalue $\lambda = A + a^2\omega^2 - 2am\omega$. Then, setting $s = -2$,
\begin{equation}
\Rop = \Delta^2\frac{d}{d r} \left(\frac{1}{\Delta}\frac{d}{d r}\right) - V(r) ,
\end{equation}
where
\begin{equation}
V(r) = V = -\frac{K^2 + 4i(r-M)K}{\Delta} + 8i\omega r + \lambda .
\end{equation}
and
\begin{equation}
K = (r^2+a^2)\omega -ma .
\end{equation}

\subsubsection{Separation of Polar Dependence and the Radial Equation}

Following \cite{GA}, if we assume that the operator $\Theta$ has regular boundary conditions at $\theta = 0$ and $\theta = \pi$, then, according to Sturm-Liouville theory, the solutions $S$ to
\begin{equation}
\Theta S + AS = 0
\end{equation} 
form a complete orthogonal set on $\theta \in [0,\pi]$ such that we can perform the expansion 
\begin{equation}
\Ltran_m = \sum_{l=|m|}^\infty R_{lm}(\omega,r)S_{lm}(\omega,\theta),
\end{equation}
where we have momentarily restored the azimuthal label $m$ for clarity. In this case, we can remove the $\theta$ dependence in the radial part of the equation by integration, while requiring that the $S_{lm}$ are normalized such that
\begin{equation}\label{eq:Orthogonality}
\int_0^\pi S_{lm}S^*_{nm}\sin\theta d\theta = \delta _{nl} ,
\end{equation}
where the asterisk denotes complex conjugation. Performing the integration, we get the radial equation
\begin{equation}\label{eq:AngularIntegration}
\Rop R_{lm} = \int_0^\pi S^*_{lm} \TI \sin\theta d\theta = \TS_{lm}(\omega,r) .
\end{equation}
To simplify the notation, let us write this as
\begin{equation}\label{eq:Teukolsky}
\Rop R = T .
\end{equation}
Up to the difference of a conventional minus sign, this is the Teukolsky radial equation with a source.

\subsection{From Teukolsky to Sasaki-Nakamura}

To make (\ref{eq:Teukolsky}) more numerically tractable, and to put it in such a form that it reduces to the Regge-Wheeler equation when $a = 0$, we transform it into SN form:
\begin{equation}\label{eq:Transformation}
\Rop R = T \to \frac{d^2X}{dx^2} - {\cal F}(r)\frac{dX}{dx} - {\cal U}(r)X = I,
\end{equation}
where $x$ is the tortoise coordinate defined by 
\begin{equation}
\frac{dx}{dr} = \frac{r^2+a^2}{\Delta} .
\end{equation}
Note that the $I$ in (\ref{eq:Transformation}) is that which appears in (\ref{eq:obspsi}). This transformation of the Teukolsky radial variable $R$ into the SN variable $X$ requires the intermediate result
\begin{equation}
R = \frac{1}{\eta}\left[\left(\alpha+\frac{\beta^\prime}{\Delta}\right)\chi - \frac{\beta}{\Delta}\chi^\prime + P \right] ,
\end{equation}
where from now on primes denote radial derivatives, $P = (r^2 + a^2)^{3/2}I$, and $\chi = X\Delta / (r^2+a^2)^{1/2} \equiv fX$. The functions $\eta$, $\alpha$, and $\beta$ are defined in~\cite{ST}.

To convey the essence of the calculations without the burden of the more detailed form of the equations, we will proceed by defining several new functions along the way. These functions may implicitly depend on $r$, $s$, $\omega$, $a$, $m$, $M$, and the angular eigenvalue $A$. Our first set of definitions merely simplifies the form of the above transformation:
\begin{equation}
R = a(r, ...)\chi + b(r, ...)\chi^\prime + s_1I.
\end{equation}
This, when we exchange $\chi$ for $X$, can be written,
\begin{equation}
R = cX + dX^\prime + s_1I,
\end{equation}
where we have again defined two new functions $c$ and $d$. We also rewrite the radial operator $\Rop$ as
\begin{equation}
\Rop = e\frac{\partial^2}{\partial r^2} + g\frac{\partial}{\partial r} + h .
\end{equation}

In this way, we are able to write the Teukolsky radial equation (\ref{eq:Teukolsky}) in terms of the SN variables:
\begin{equation}\label{eq:Reduce}
\Rop R = (e\frac{\partial^2}{\partial r^2} + g\frac{\partial}{\partial r} + h)(cX + dX^\prime + s_1I) = T.
\end{equation}
This equation contains a third-order derivative, while the SN equation has at most a second-order derivative. Our next step then is to reduce the order of (\ref{eq:Reduce}). We will do this by differentiating the SN equation to find the third-order derivative in terms of lower orders, and substituting the result into the above. This will allow us to write the SN equation with an explicit source term derived from the initial conditions.

\subsubsection{Reducing the Third-Order Derivative}

The SN equation with a source,
\begin{equation}\label{eq:SN}
\frac{d^2X}{dx^2} - {\cal F}(r)\frac{dX}{dx} - {\cal U}(r)X = I,
\end{equation}
is written in terms of the tortoise coordinate and derivative. In (\ref{eq:Reduce}), the derivatives are radial. To relate them, we must change coordinates from $x$ to $r$. Applying the chain rule, we obtain
\begin{equation}
\frac{\Delta}{r^2+a^2}\frac{d}{dr}\left(\frac{\Delta}{r^2+a^2}\frac{dX}{dr}\right) - \frac{\Delta}{r^2+a^2}{\cal F}(r)\frac{dX}{dr} - {\cal U}(r)X = I.
\end{equation}
It will help us to use the form of this equation where the second-order derivative term is isolated:
\begin{equation}
\frac{d^2X}{dr^2} = \frac{(r^2+a^2)^2}{\Delta^2}\left( \left(\frac{\Delta}{r^2+a^2}{\cal F}(r) - \frac{\Delta}{r^2+a^2}\frac{d}{dr}\left(\frac{\Delta}{r^2+a^2}\right)\right)\frac{dX}{dr} + {\cal U}(r)X +I\right).
\end{equation}
Again, for simplicity, we can define new functions such that
\begin{equation}
X^{\prime\prime} = FX^\prime + UX + HI.
\end{equation}
To get the third-order derivative in terms of lower orders, we differentiate:
\begin{equation}
X^{\prime\prime\prime} = F^\prime X^\prime + FX^{\prime\prime} + U^\prime X + UX^\prime + H^\prime I + HI^{\prime}.
\end{equation}
Or
\begin{equation}\label{eq:Third}
X^{\prime\prime\prime} = FX^{\prime\prime} + JX^\prime + U^\prime X + HI^{\prime} + H^\prime I.
\end{equation}

Expanding (\ref{eq:Reduce}) we get
\begin{equation}
(ed)X^{\prime\prime\prime} + (ec+2ed^\prime+gd)X^{\prime\prime} + (2ec^\prime+ed^{\prime\prime}+gc+gd^\prime+hd)X^\prime + (ec^{\prime\prime}+gc^\prime+hc)X = T - \Rop (s_1I).
\end{equation}
Substituting in (\ref{eq:Third}) and collecting the 0th power SN variable terms on the source side, we obtain
\begin{equation}
\begin{split}
& (edF+ec+2ed^\prime+gd)X^{\prime\prime} + (edJ+2ec^\prime+ed^{\prime\prime}+gc+gd^\prime+hd)X^\prime + (edU^\prime+ec^{\prime\prime}+gc^\prime+hc)X \\
& = T - edHI^\prime - edH^\prime I  - \Rop (s_1I).
\end{split}
\end{equation}

\subsubsection{The SN Equation with a Source}

We now convert derivatives on the left hand side back to tortoise coordinate derivatives which requires an application of $dx/dr = (r^2+a^2)/\Delta$ for each prime: 
\begin{equation}
X^{\prime\prime} = \frac{(r^2+a^2)^2}{\Delta^2}\frac{\partial^2 X}{\partial x^2} + \frac{d}{dr}\left(\frac{r^2+a^2}{\Delta}\right)\frac{dX}{dx},\quad X^\prime = \frac{r^2+a^2}{\Delta}\frac{dX}{dx}
\end{equation}
So we have
\begin{equation}
\begin{split}
& \frac{(r^2+a^2)^2}{\Delta^2}(edF+ec+2ed^\prime+gd)\frac{\partial^2 X}{\partial x^2} + \\
& \left(\frac{d}{dr}\left(\frac{r^2+a^2}{\Delta}\right)(edF+ec+2ed^\prime+gd) + \frac{(r^2+a^2)}{\Delta}(edJ+2ec^\prime+ed^{\prime\prime}+gc+gd^\prime+hd) \right)\frac{\partial X}{\partial x} + \\
& (edU^\prime+ec^{\prime\prime}+gc^\prime+hc)X = T - edHI^\prime - edH^\prime I  - \Rop (s_1I) ,
\end{split}
\end{equation}
which, with some relabelling, we can write as
\begin{equation}
E\frac{\partial^2 X}{\partial x^2} + B\frac{\partial X}{\partial x} + CX = T - DI - PI^\prime  - \Rop (s_1I),
\end{equation}
in which case it is easy to extract the SN equation with the source:
\begin{equation}
\frac{\partial^2 X}{\partial x^2} + \frac{B}{E}\frac{\partial X}{\partial x} + \frac{C}{E}X = \frac{T-DI-PI^\prime  - \Rop (s_1I)}{E} .
\end{equation}
We now have the source in terms of the original initial conditions set in the Teukolsky framework. According to our definitions, we identify that our functions are related to the SN functions by
\begin{equation}\label{eq:Comparison}
\frac{B}{E} = -{\cal{F}},\quad \frac{C}{E} = -{\cal{U}},\quad \frac{T-DI-PI^\prime  - \Rop (s_1I)}{E} = I ,
\end{equation}
and at this stage it is possible to numerically solve the last of these relations for the desired source, $I$.

\section{The Modulated Transfer Function and Echoes}

Now that we have derived the source term, we may see what its affect is on the transfer function, and in the process generate a more realistic echo pattern. We will begin by making clear certain steps of the procedure where assumptions and approximations must be made. For definiteness, we will work with the example of a Gaussian initial perturbation, and set background parameters to correspond to values typical of the gravitational wave events so far observed by the LIGO and VIRGO collaborations.

\subsection{Settings, Approximations, and Assumptions}

To gain numerical results from our source function, we will be required to make choices in two places. First is in the separation of the $\theta$ dependence, where integration and the orthogonality of the eigenfunctions were used. Second is in the nature of the initial perturbation, manifested in the initial condition functions $u$ and $v$.

\subsubsection{Separation of the $\theta$ Variable}

To derive the radial Teukolsky equation with a source, we multiplied by eigenfunctions of the $\Theta$ operator and used their orthogonality to eliminate $\theta$-dependent terms. To obtain the radial source term $T$, which is given by (\ref{eq:InitialConditions},\ref{eq:AngularIntegration},\ref{eq:Teukolsky}), the relevant integrals are
\begin{equation}
I_1 = \int_0^\pi S_{lm}S^*_{nm}\sin^3\theta d\theta, \quad I_2 = \int_0^\pi S_{lm}S^*_{nm}\sin\theta\cos\theta d\theta,
\end{equation}
along with the orthonormality relation (\ref{eq:Orthogonality}). The eigenfunctions $S_{lm}$ are spin-weighted spheroidal harmonics, and to evaluate these integrals we use the Black Hole Perturbation Toolkit~\cite{Warburton:BHPT}. 

For all merger events which have so far been observed, the remnant objects have spins close to $a = 0.7M$. Since spin-weighted spheroidal harmonics are functions of the combination $a\omega$, we can choose a particular value of $a$ and write them as functions of frequency. Choosing $a=0.7M$ to match the current data, the resulting integrals depend nearly linearly on frequency, up to values far in excess of the primary ringdown frequency. In this case, to a very good approximation, $I_1 \approx -0.070\omega + 0.47$ and $I_2 \approx 0.062\omega + 0.67$. Again from the Toolkit, while still setting $a=0.7M$, we also find the good approximation $\lambda \approx -4.5\omega + 3.9$.

\subsubsection{The Initial Condition Functions $u$ and $v$: Gaussian Example}

Here, for definiteness, we make the physically reasonable assumption that the initial perturbation is Gaussian in the radial tortoise coordinate and spheroidally harmonic in the polar and azimuthal angles in such a way that the dominant ($l=m=2$) mode is primarily supplied while others can be ignored. Under these assumptions, the initial conditions become
\begin{equation}
\begin{split}
& u = \psi_m(t,x,\theta)|_{t_0} \approx g(t,x)|_{t_0}S_{22} = e^{-(t_0+x-x_0)^2/w^2}S_{22}, \quad \\
& v = \partial_t \psi_m(t,x,\theta)|_{t_0} \approx \partial_t g(t,x)|_{t_0}S_{22} = \frac{-2(t_0+x-x_0)}{w^2}e^{-(t_0+x-x_0)^2/w^2}S_{22},
\end{split}
\end{equation}
where the starting time $t_0$ can be taken to be 0, and $x_0$ and $w$ set the initial location and width of the pulse. Proceeding thus, we can apply these assumptions to the radial equation
\begin{equation}
\Rop R = \int_0^\pi S^*_{lm} \TI \sin\theta d\theta = T,
\end{equation}
in which case we obtain
\begin{equation}
\begin{split}
T = & - \left(-i\omega\left[\frac{(r^2+a^2)^2}{\Delta}-a^2I_1\right] + \frac{4iMmar}{\Delta} - 2s\left[\frac{M(r^2-a^2)}{\Delta} - r - iaI_2\right]\right)e^{-(x-x_0)^2/w^2}\\
& + \left(\frac{(r^2+a^2)^2}{\Delta}-a^2I_1\right)\frac{2(x-x_0)}{w^2}e^{-(x-x_0)^2/w^2} .
\end{split}
\end{equation}
Or, combining terms,
\begin{equation}
\begin{split}
T = & - \left(-i\omega\left[\frac{(r^2+a^2)^2}{\Delta}-a^2I_1\right] + \frac{4iMmar}{\Delta} - \right. \\
& \left. 2s\left[\frac{M(r^2-a^2)}{\Delta} - r - iaI_2\right] - \frac{2(x-x_0)}{w^2}\left[\frac{(r^2+a^2)^2}{\Delta}-a^2I_1\right]\right)e^{-(x-x_0)^2/w^2} .
\end{split}
\end{equation}
In the above we have written the equations in terms of both $r$ and $x$, which is possible since there is a one-to-one correspondence for $x=x(r)$. At this point, we have fully specified all functions needed to calculate the SN source (\ref{eq:Comparison}).

\subsection{Signal Modulation}\label{subsec:Modulation}

To see the implications of including a source function for data searches, we return to (\ref{eq:obspsi}) which gives the observed spectrum for echoes, $\psi_\omega$. It contains two parts: 1) The transfer function ${\cal K}$, and 2) The source integral which modulates it.

\begin{figure}[h]
\centering
\includegraphics[width=0.7\textwidth]{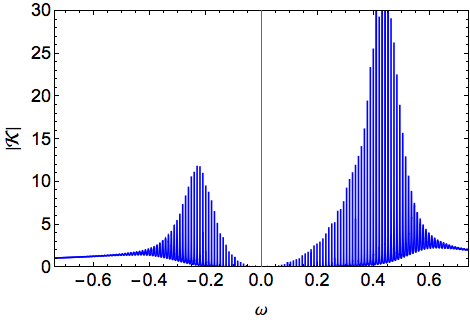}
\caption{\label{fig:KT1} The transfer function, a series of resonances extending to positive and negative frequencies which is asymptotically approaching unity in both directions. The spacing of the resonances is roughly equal to the inverse of twice the cavity size, and their locations depend on the orbital numbers and BH parameters. Here, we have chosen a wall at $x_\textrm{wall}=-300$, $a=0.7M$, and $l=m=2$. The boundary condition used to generate this plot corresponds to perfect energy reflection at the wall.}
\end{figure}

As has been described in~\cite{Windows,AnalyticalEchoes}, ${\cal K}$ is characterized by a series of nearly evenly spaced resonances. It depends on the normalization of the homogenous solutions of the perturbation equation, but any normalization which is only a function of frequency can be chosen without affecting the observed spectrum. The reason for this is that ${\cal K} = 1/W$, where $W$ is the Wronskian of the system and proportional to the normalization, so the normalization appears in the denominator of (\ref{eq:obspsi}), but the normalization also appears in the numerator via $\psi_\textrm{left}$. We will choose a frequency-independent normalization to best display the affect of the source.\footnote{In~\cite{Windows}, the transfer function is defined with an extra normalization factor of $\sqrt{k_H\omega}$ compared to the normalization here. In~\cite{AnalyticalEchoes}, the normalization depends on the spectral energy density.} A sample ${\cal K}$ is given in Fig.\ref{fig:KT1}.

\begin{figure}[h]
\centering
\includegraphics[width=1\textwidth]{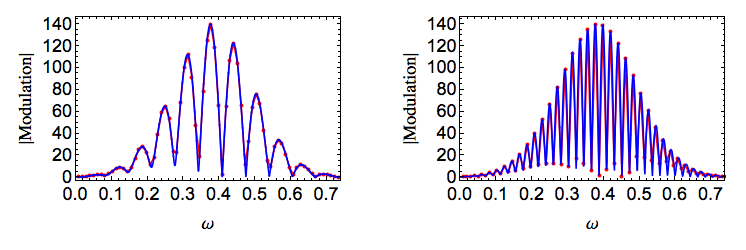}
\caption{\label{fig:Modulation2} Left: The modulation integral for a Gaussian of width $w=12$ at $x_0=-250$, near the wall at $x_\textrm{wall}=-300$. Right: The same but for $x_0=-150$. The blue curves are the absolute values of each modulation integral, while the red dots are the values at the resonance frequencies. These curves were generated for a truncated Kerr background of spin $a=0.7M$. The transfer function, importantly, contains negative frequency modes as well, but for this example the modulation practically kills everything outside the plotted region. Other types of sources can support negative modes, in which case a wider region must be examined.}
\end{figure}

\begin{figure}[h]
\centering
\includegraphics[width=1\textwidth]{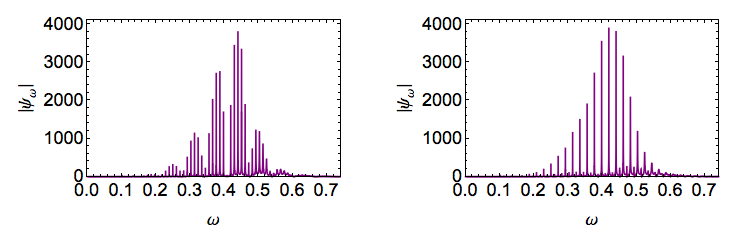}
\caption{\label{fig:Observation2} Left: $\psi_\omega$ for a Gaussian of width $w=12$ at $x_0=-250$, near the wall at $x_\textrm{wall}=-300$. Right: The same but for $x_0=-150$. The observable $\psi_\omega$ is the product of the transfer function ${\cal K}$ and the source integral. Because ${\cal K}$ is dominated by resonances, the affect of the source is essentially a modulation of peak heights. The amount by which each peak is modulated is shown by the red dots in Fig.\ref{fig:Modulation2}. While in each case the observational signal has significantly deviated from Fig.\ref{fig:KT1}, the second displays an interesting pattern where every second peak has nearly been deleted. Such a signal observed in the data could lead to incorrect conclusions about the geometry of the system.}
\end{figure}

Since ${\cal K}$ is dominated by the characteristic resonances, we can determine how it is modulated by evaluating the integral at the resonance frequencies. We will primarily be interested in the magnitude of the frequency content of the signal, which corresponds to the absolute value of the observable $\psi_{\omega}$. In Fig.\ref{fig:Modulation2}, we plot the magnitude of the source integrals (blue curves) and their values at the resonance frequencies (red dots) for Gaussian initial perturbations placed at different locations in the cavity. The affect of this modulation on the transfer function is shown in Fig.\ref{fig:Observation2}. In both cases the pattern of evenly spaced resonances is significantly distorted. For the Gaussian in the centre of the cavity, every second peak is practically deleted. In a real data search, such a modulation could lead to a deceptive time delay and cavity size estimation  (Fig.\ref{fig:Echoes3}). For other types of simple initial conditions, many such deceptive patterns can be generated  (Fig.\ref{fig:EchoesD}).

The shape of these blue curves is due to the size and location of the initial Gaussian perturbation. Within the cavity, the solutions to the homogenous SN equation, $\psi_\textrm{left}$, are composed of complex exponentials of the form $e^{\pm i k_H (x-x_\textrm{wall})}$. Additionally, in the same region, the p-function takes on constant values, and so the modulation integrand is essentially the source multiplied by sinusoidal waves. 

The frequencies of these waves are centred around the Kerr frequency related to the spin of the background, $k_H = \omega - m(a/2r_+)$, where $r_+$ is the outer horizon radius. For $M=1$, $a=0.7$, and $l=m=2$, the Kerr frequency becomes $k_H = \omega - 0.408$. When $\omega = 0.408$, the homogenous solutions on the inside have infinite wavelength and the integrand is the source $I$ multiplied by a constant. As frequencies significantly deviate from $\omega = 0.408$, a large enough Gaussian covers several full periods of the homogenous solution, and the net result is small, hence the drop in magnitude of the blue curves away from the central frequency. 

\begin{figure}[h]
\centering
\includegraphics[width=1\textwidth]{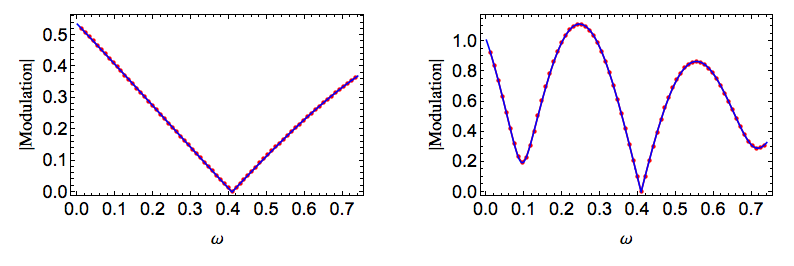}
\caption{\label{fig:Delta} A delta function source in a cavity with $x_\textrm{wall} = -300$. Left: Modulation for a delta function located very close to the wall with $x_0 = -299$. Right: Modulation for a delta function located slightly further away with $x_0 = -290$. Were such sources to be present astrophysically, the affect on the signal would be severe in that large groups of resonances could be removed.}
\end{figure}

The individual oscillations are due to the phase of the part of $\psi_\textrm{left}$ covered by the Gaussian. For example, a Gaussian the same size as one period of $\psi_\textrm{left}$ will yield a value of about $0$ for the integral when centred on the midpoint of a wave. On the other hand, in the limiting case of a Gaussian of negligible width, (i.e. a delta function) the source picks up the value $\psi_\textrm{left}$ takes at $x_0$. Therefore, a narrow Gaussian can contribute significantly even as $\omega \to 0$, where the effective frequency is higher. This is shown in Fig.\ref{fig:Delta} for delta functions placed near the wall.

As suggested by these results, any number of modulation curves may be generated for various initial conditions. For example, if we were to place a Guassian near the potential, that is $x_0 \approx 0$, oscillations of the blue curve would be separated by nearly the same spacing as resonances of the transfer function, and red dots would at first be located near peaks of the blue curves. However, the slight offset of the Gaussian from the potential implies slightly unequal spacing between blue peaks and red dots, an affect which would accumulate with changing frequency, the result of which would be the modulation featuring its own frequency which would subdivide the transfer function into smaller groups of resonances. Many additional examples could be discussed.

\subsection{Echoes in the Time Domain}

\begin{figure}[h]
\centering
\includegraphics[width=0.7\textwidth]{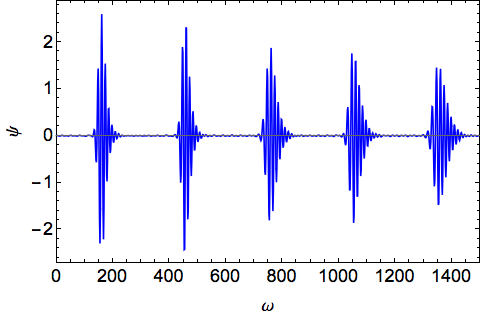}
\caption{\label{fig:Echoes3} Echoes generated from a Gaussian initial perturbation with an outgoing velocity located in the centre of the cavity with $x_0 = -150$. Nontrivial frequency, amplitude, and phase information is carried by each echo. The essential information can be extracted more clearly from the modulated transfer function in frequency space. Though the cavity size corresponds to $x_\textrm{wall}=-300$, the time delay here is $t_\textrm{d} \approx 300$ instead of the usual $t_\textrm{d} \approx 2|x_0|$; this is the affect of every second resonance being deleted in the spectrum.}
\end{figure}

While echoes are in many cases most effectively studied in the frequency domain, it is nevertheless interesting to examine their behaviour in time. We can do this by applying the inverse Laplace transformation according to (\ref{eq:InverseLaplace}). In Fig.\ref{fig:Echoes3}, we plot the first several of a train of echoes. 

The pattern generated possesses several characteristics which are not features of some popular templates currently in use. Notably, the frequency content of subsequent echoes is changing, and the amplitudes are not obviously gained by application of some power of a large constant factor.

\begin{figure}[h]
\centering
\includegraphics[width=0.7\textwidth]{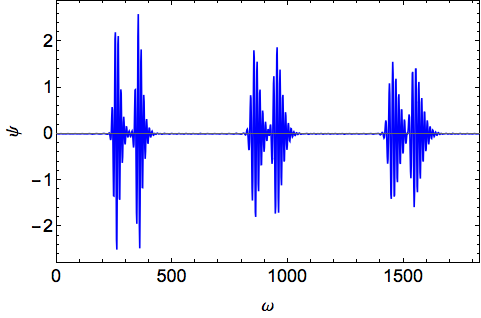}
\caption{\label{fig:EchoesD} The same example as in Fig.\ref{fig:Echoes3}, but with $x_0=-250$. The delay between adjacent pulses is about $100$ while the time delay between double-pulses is closer to the typical expectation of $600$. This pattern already significantly deviates from the basic template since it features two time delays. The situation can be arbitrarily complicated by adding additional initial pulses in various configurations. The general form of the characteristic spectrum is somewhat robust against such changes.}
\end{figure}

The echoes embody significant structure which would be difficult to model precisely in a matched-filtering method, the phase information of each echo being one example. In this way, it may be more effective to use frequency-based methods which isolate information from the characteristic resonances. For example, by taking the absolute value of $\psi_\omega$ and working in the frequency domain, phases are ignored but information on the time delay is clearly perceived. Moreover, the dependence of the echo pattern on initial conditions is significant. For the same initial perturbation, but with the pulse located near the wall instead of in the centre of the cavity, a double-pulse echo pattern is observed which we display in Fig.\ref{fig:EchoesD}

Either way of approaching echoes leads to difficulties, and it may be that, at present, one must choose carefully which properties to highlight, at least until noise reduction in detectors is sufficiently improved to begin deeper analysis. 

\section{Conclusions}

We have calculated the source term for gravitational perturbations on a Kerr background starting from initial conditions. The initial conditions were originally defined in the Teukolsky framework and then transformed into Sasaki-Nakamura form, and the result was the usual SN equation but with an explicit source term given by (\ref{eq:Comparison}).

The source term was then used in (\ref{eq:obspsi}) to calculate its modulating affect on the transfer function, a series of resonances which is a signal target in echo searches. This provided several insights as to how the characteristic resonance pattern for echoes may change for different source locations and shapes, and we discussed implications for data analysis. Finally, we inverted the frequency space results to obtain physically motivated echo patterns, which turned out to be quite distinct from the typical templates currently in use.

\section*{Acknowledgements}

I would like to thank Bob Holdom for several helpful discussions.

\end{document}